# Bending Energy-Driven Cooperative Patterning of 2D Colloids in Elastic 2D Fluids

Weiyue Xin,[1] and Maria M. Santore[2]*

1. Department of Chemical Engineering
   University of Massachusetts Amherst

2. Department of Polymer Science and Engineering
   University of Massachusetts Amherst
   120 Governors Drive
   Amherst, MA 01003
   santore@mail.pse.umass.edu

## Abstract

Suspensions of colloidal microplates in contoured 2D elastic fluids sheets are dominated by the bending mechanics and shear rigidity of the plates and the contrasting in-plane shear flow of the 2D fluid. Using the phase separated phospholipid *membranes* of individual giant unilamellar vesicles as models of contoured 2D suspensions, where solid domains act as colloids in a fluid membrane, we explore bending elasticity-driven assembly. The plate-shaped domains are varied between 1-10 μm in diameter, with 4-100 plates per vesicle depending on size, contributing a solid area of 17 ±3%. Three classes of reversible plate arrangements evidence inter-plate attractions and repulsions: persistent hexagonal vesicle-encompassing quasi-lattices, persistent closely associated configurations (chains or concentrated lattices), and a dynamic disordered state. The vesicle-encompassing quasi-lattice is stable to vesicle dehydration by 30% relative to an inflated sphere. Excess area or membrane slack, for a fixed composition, dominates the preferred configuration while domain size and number contribute pattern intricacy. Different from the gradual variations in domain interactions and tunable positions in two-colloid systems, multibody interactions vary sharply within a particular range of excess area, producing cooperative assembly reminiscent of a phase transition.

Significance Statement:
The ability of elastic interactions to produce patterns in membrane objects, and the sharp sensitivity patterning development relative to disorder upon the change in membrane flaccidness, suggests a phase transition which could be exploited in various applications and which provides insights into biological behaviors. Sharp changes in areal density and in quasi-2D structure may enable new fabrication methods for thin patterned thin films, or could form the basis for ultra-thin composites with responsive properties: optics, electrical connectivity, transport behavior, or mechanics. In biological systems, the sharp structural change of the arrangements of membrane objects, occurring with differences in membrane flaccidity, suggests a route to cellular function.

Classification: Physical Sciences: Applied Physical Sciences, Engineering

**Introduction**

Ultra-thin bendy composite sheets, dominated by properties in their lateral directions and conceptualized as being two-dimensional but contoured, are a new class of materials imbued with mechanical response and reconfigurability. Such 2D materials may some day form the basis for color changing skins, touch sensors, and reassembling circuit-containing films. Despite their novelty, 2D liquid-solid composite sheets have long existed in the form of cell membranes.[1] These fluid phospholipid bilayers change shape and direct the assembly of molecules, lipid rafts, and nano-scale objects such as capsids.[2-5]

Studies targeting an understanding of living cells often exploit giant unilamellar phospholipid vesicles to capture the essential membrane physics. These efforts have revealed how functional interactions are regulated by the membrane curvature produced by embedded or adhered objects, macromolecules, and lipid rafts themelves.[6] For instance, though line tension drives the coalescence of discrete fluid membrane domains,[7,8] line tension can also distort fluid domains to produce stabilizing interdomain repulsions.[9-11] This line-tension driven instability, causing domains to bulge, occurs only when domains are large enough and vesicles sufficiently under-inflated. The resulting long-range repulsion can give rise to domain ordering, for instance producing raspberry-like lattices of bulging domains.[10]

In contrast with the repulsive interactions of fluid domains, adhesive nanoparticles[12-14] and membrane proteins[6,15-17] act as attractive inclusions. Beyond the physico-chemical influence from molecular disruption near the point of adhesion, the membrane curvature resulting from partial engulfment produces longer range attractions that concentrate small objects.[6,18] Adherent nanoparticles can form plaques with closely associated hexagonal order or belts of more distantly-associated inclusions that encircle the waist of underinflated vesicles.[12] A competition between adhesion and membrane bending has been suggested to govern the regimes where these different structures are seen. Separately, simulations of NBar proteins reveal string-like structures where protein features arrange regularly along the strings, implicating the detailed protein structure in the assembly morphology.[19]

Flat membrane objects comprise a different type of inclusion. In this category, solid phosphatidylcholine membrane domains possess molecular order and symmetry across the their bilayers,[20-23] tending to be flat[24,25] and rigid against shear.[25,26] Thus, discrete solid domains behave as plate-shaped colloids within otherwise fluid phospholipid membranes. In these 2D fluid-solid composites, both the solid plates (2D particles) and the fluid membrane can bend and undergo limited stretching. However, unlike the plates, the fluid can shear in-plane, allowing it to take on Gaussian curvature.[26] We recently demonstrated how the resulting pairwise inter-plate interactions, with two in each vesicle membrane, contained both attractions and repulsions that enabled the plates to rest at a fixed separation with only submicron fluctuations in their relative positions along the membrane contour.[24] The excess membrane area, defined as the actual vesicle area normalized on the area of a sphere of equal volume, was the key parameter in determining the relative plate positions in these 2-colloid systems. Further by tuning excess area through osmotic inflation or micropipette manipulation, the domains could be reversibly repositioned between relatively stable states. A simple model demonstrated that this behavior could be reproduced with fluid bending as they key material property and not requiring a line tension or physico-chemical interactions of the domains. The treatment predicted that the impact of excess area would decrease with small domain sizes or area fractions. Thus is was established that plate shaped colloids experience tunable mechanical interactions that differ from exclusively fluid domains or nanometric inclusions in a membrane.

The current work examines multibody assemblies in 2D elastic suspensions of colloidal plates and highlights morphologies that cannot be anticipated from the current understanding of pairwise interactions. Directed by fluid membrane bending, but different from the instability-induced repulsive lattices of fluid domains, the plates appear to maintain zero Gaussian curvature while arranging into ordered lattices, chains, or ordered plaques suggestive of combined attractions and repulsions sometimes with anisotropic character. This variety of structures is explored for systems near a single composition, with a 17 ±3% solid area fraction. Variations in colloid size and number change, over an order of magnitude, the intricacy but not the underlying character of the assembled patterns. We introduce the excess area, which regulates the bending energy of the system and dominates the state space, switching interactions from attractive to repulsive. In contrast with the gradual shifting seen with two colloidal plates,[24] the multi-plate systems assemble sharply and cooperatively with a small difference in excess area. Thus a variety assembled configurations, from repulsive lattices to chains of associated plates, can be formed using a single material composition.

**Results.**

Compositionally uniform giant unilamellar vesicles were produced by electroformation from films of lipid mixtures at elevated temperatures in the one phase region of the phase diagram. After room temperature storage, vesicle suspensions were reheated and cooled at controlled rates to produce vesicle membranes containing relatively monodisperse compact solid domains. With the solid domains visualized by their exclusion of a fluorescent tracer lipid, the solid area fraction, $\phi$, measured from fluorescence micrographs was found to be nearly the same for all vesicles, $0.17 \pm 0.03$. For an overall lipid composition of 40 mol% DPPC (1,2-dipalmitoyl-sn-glycero-3-phosphocholine), 60 mol% DOPC (1,2-dioleoyl-*sn*-glycero-3-phosphocholine) , calculations based on the phase diagram and physical property data, in the Supporting Information, predict a solid area fraction consistent with these observations. At room temperature, the solid domains were stable against coalescence, maintained a fixed shape, and appeared flat or nearly so, consistent with prior work.[24] For vesicles 10-40 μm in diameter, the numbers of domains per vesicle, 4-100, was controlled via the cooling rate. With a fixed solid area fraction set by the overall lipid composition, the domain number $N$ per vesicle of diameter $D_v$ is related to domain size via

$$D_d/D_v = \sqrt{4\phi/N}.$$ \hfill (1)

In this work $D_d/D_v$ varied in the range 0.05 - 0.45. Because the solid micron-scale domains are thin, flat, and exhibit Brownian motion when not strongly interacting, they behave as two dimensional colloidal particles. We therefore refer to them as *colloidal plates*.

*Arrangements of Colloids in Membranes*

Figure 1 presents examples and summarizes features of the 3 types of patterns of colloids observed in these systems: persistent vesicle-encompassing hexagonal quasi-lattices, persistent closely associated configurations (chains or concentrated lattices), and a dynamic disordered state. In the closely associated state, domains appeared not to touch. Configurations were classified as persistent if the colloidal plates visually exhibited limited translation on the timescale of minutes. Persistent assemblies were not, however, permanent and could be disrupted with gentle touch, later reforming similar but not identical structures. In the dynamic disordered state, domains translated on length scales of their own diameters within several minutes. While the insets of each micrograph show the Delaunay triangulation analysis discussed below, close-ups are included in the Supporting Information.

The vesicles exhibiting any of three classifications of colloid arrangements could be found within single electroforming and/or reheating batches, and we observed no obvious influence of vesicle size on the category of pattern found. Osmotic conditions, however, were found to be important in determining which category of pattern was observed. Figure 1D, summarizes the distribution of patterns found at neutral osmotic conditions, 100 mOsm both inside and outside vesicles. Here vesicles of many sizes presented vesicle-encompassing hexagonal quasi-lattices or the dynamic disordered state. By contrast in Figure 1E, when the vesicle suspension was diluted 50% in DI water so that the external osmolarity was reduced to 50 mOsm, conditions which tended to inflate vesicles, the distribution of patterns was shifted to favor closely associated configurations or the dynamic disordered state. Thus vesicles of the same material composition and batch were osmotically configured, to shift colloidal interactions and resulting assemblies.

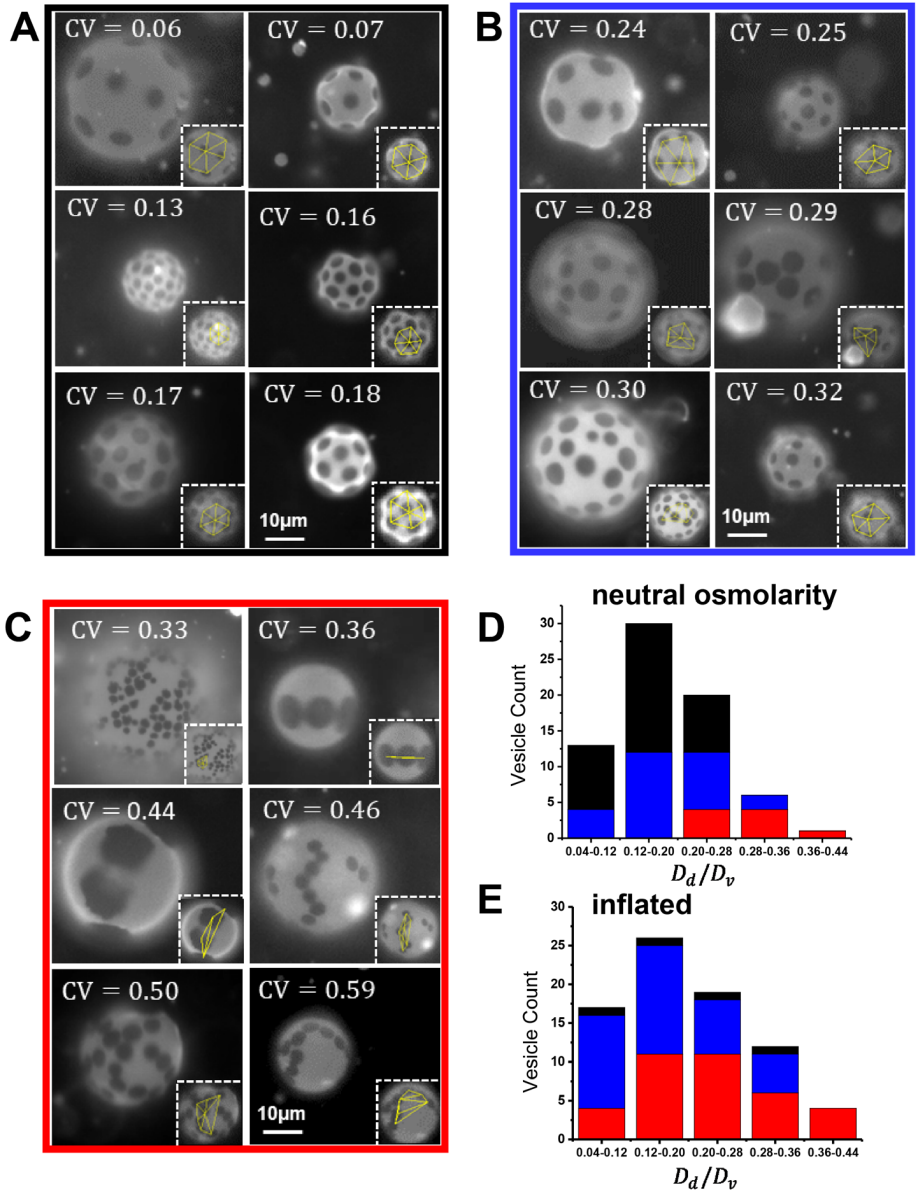

**Figure 1**. Examples of 6 vesicles for each of three pattern classifications: A) Vesicle-encompassing hexagonal quasi-lattice (black); B) Dynamic disordered state (blue); and C) Closely associated configurations (red). Each inset shows the limited Delaunay triangulation, featured at larger magnification in the SI. The histograms in D and E show the distributions of different pattern classifications found at D) neutral (100 mOsm inside and outside) and E) inflated (100 mOsm inside and 50 mOsm outside) osmotic conditions.

While the types of arrangements are discernable by eye, their differences were also systematically distinguished via image analysis. We employed a limited Delaunay triangulation method, accounting for the fact that not all plates could be in focus at once, detailed in the Supporting Information. Focusing on the top or bottom of a vesicle, the most central plate and up to 6 neighbors were identified. Their centers were connected by Delaunay triangulation, as shown in the inset of each image of Figures 1A-C. The lengths of the center-center lines were averaged, and the standard deviation, divided by this average defines the Coefficient of Variation, $CV$.

In the examples in Figures 1A-C, and for 151 vesicles in total, the $CV$ values for each vesicle fell into distinct clusters, consistent with the initial visual assessment of pattern appearance. For instance for the vesicle-encompassing quasi-lattice configurations, $CV < 0.2$; for the vesicle with disordered dynamic positions, $0.23 < CV < 0.33$; and for the closely associated structures, $CV > 0.34$. Therefore the limited Delaunay approach to pattern classification, which upholds the by-eye classification, established a lack of observer bias.

*Vesicle Inflation and Excess Membrane Area*

The impact of osmotic conditioning to inflate vesicles and shift the distribution of domain patterning in populations of vesicles in Figure 1D, motivates quantification of the excess area $A_{xs}$. $A_{xs}$ is defined as the vesicle membrane area normalized on the area a sphere of equal volume.[24,27] Therefore

$$A_{xs} = \frac{A}{(4\pi)^{\frac{1}{3}} * (3V)^{\frac{2}{3}}} \qquad (2)$$

While $A_{xs}$ is related to reduced volume employed in other works,[28-30] $v_r = A_{xs}^{-3/2}$, focusing on $A_{xs}$ emphasizes physical meaning at the level of the membrane. $A_{xs}$ is measured, as described in the Supporting Information, by aspirating vesicles at low suctions into micropipettes, enabling both their volume and surface area to be quantified. Images for low-suction aspiration of vesicles exhibiting the three classes of colloid arrangements are summarized in Figure 2A-C, along with, in Figure 2D, a summary plot illustrating the relationship between excess area and the Coefficient of Variation, $CV$, of the colloid arrangement prior to aspiration. This plot includes 143 vesicles studied by image analysis and then by micropipette aspiration, from 7 different electroforming batches, and 15 reconditioning runs. A strong correlation of $A_{xs}$ and colloid pattern reveals the highest $A_{xs}$ for the vesicle encompassing quasi-lattices and the smallest $A_{xs}$ values for associated chains and ordered clusters.

Also worthy of comment, while we were not able to quantitively analyze entire vesicles, different focal plans through vesicles exhibiting vesicle-encompassing quasi-lattices suggest complete patterning of the entire vesicle. The nearly hexagonal arrangements of domains and their regular spacing is visually striking, especially given the variety in domain numbers per vesicle. Close scrutinization does reveal limited numbers of domains with five rather than six neighbors, as would be expected for patterning on a sphere. The term quasi-lattice is employed here for rigor since it is not possible to hexagonally pattern a sphere. Nonetheless, it is clear that domains approximate a hexagonal lattice to a great extent.

The Lindemann parameter provides a measure of the fluctuations of objects in lattices relative to their mean spacing, with $LP = CV/\sqrt{2}$.[31] Since fluctuations grow large towards the melting point, a threshold value of the LP is often invoked as a criterion for equilibrium melting to a fluid state.  Threshold values fall in the range 0.05-0.2, depending factors like interaction type and dimensionality. While there are complexities in applying the Lindemann melting criterion in 2D systems,[32,33] even discounting the effects of finite size and spherical curvature, we note that in Figure 2D the $CV$ resulting from the limited triangulation method suggests a melting transition at a Lindemann parameter of 0.14 or $CV = 0.2$ between the vesicle encompassing quasi-lattice configuration and the dynamic disordered state.  A full analysis taking into account all colloidal plates on each vesicle, beyond our imaging capabilities, is required for rigorous quantitative analysis of vesicle-encompassing quasi-lattice structure in the future.  None the less, the boundary between the vesicle encompassing quasi-lattice configuration and the dynamic disordered state is, from the perspective of the Lindemann criterion, suggestive of a melting transition.

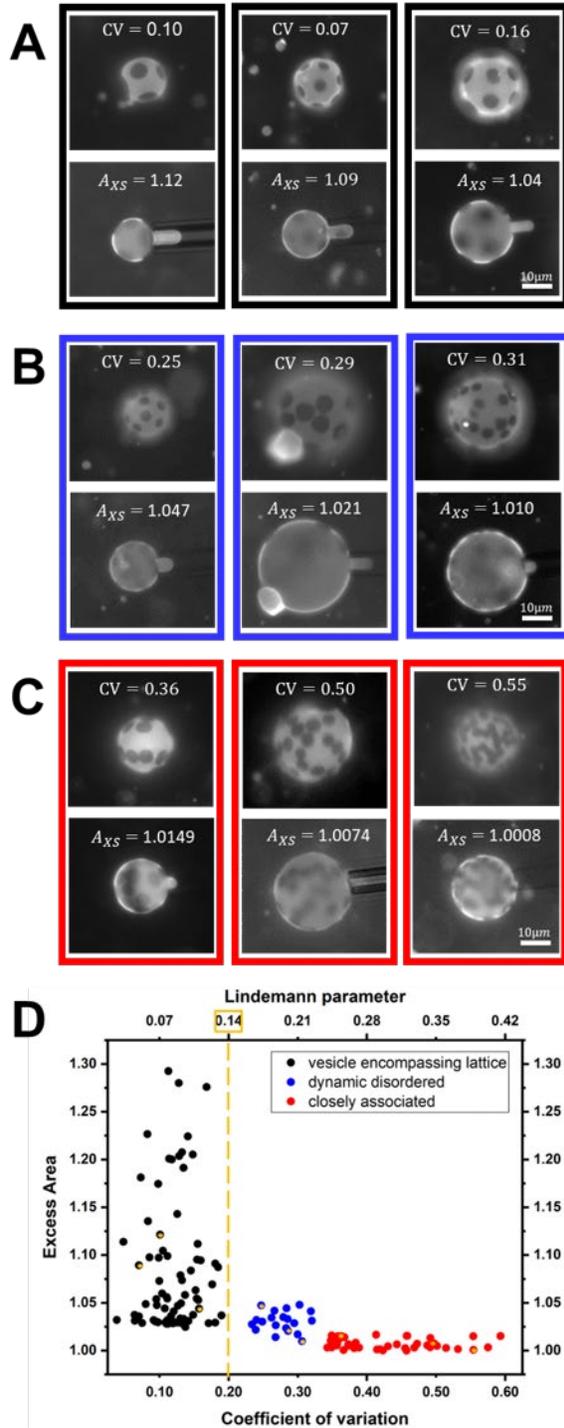

**Figure 2**. Examples of micropipette aspiration of vesicles at low suction to determine excess area. Three vesicles are shown for each of the different pattern types: A) Vesicle-encompassing quasi-lattice; B) Dynamic disordered state; and C) Closely associated configurations. The images show the greatest membranes projections into the pipette for the vesicle-encompassing lattice pattern. The scale bars apply to all images. D) The relationship between the excess area and the coefficient of variation from the limited Delaunay triangulation measured for each vesicle. Vesicles featured in A-C are starred in D.

*State Space for Colloidal Patterns*

A potential correlation between excess area and pattern type in Figure 2 motivates the state space of  Figure 3, with $A_{xs}$ and the number of colloidal plates per vesicle, $N$, as the main parameters. When plates are uniformly distributed, as in the vesicle-encompassing quasi-lattice, in addition to the relationship in Equation 1, $N$ also relates to an approximate center-center domain spacing, $D_{cc}$, normalized on the vesicle diameter, $\frac{D_{cc}}{D_v} = \sqrt{4/N}$. This relationship is confirmed in the Supporting Information, for 78 vesicles that are fully encompassed by domains in a quasi-lattice configuration.  The relationship persists with moderate vesicle deflation due to similar reductions in $D_{cc}$ and $D_v$ when the membrane exhibits bending fluctuations.

Figure 3A reveals that data points broadly cover the region of state space for the vesicle-encompassing quasi lattice configuration, indicating this configuration is observed over a large range of values in $A_{xs}$, $N$, and $D_{cc}/D_v$.  Quite distinct, however, is the sharp boundary at $A_{xs} \sim 1.02$ below which no vesicle-encompassing quasi-lattice configurations were found.  Figure 3B shows example quasi-lattice structures on vesicles along the lower boundary, as the domain size is increased, or $N$ is decreased at constant solid area fraction, *over 1.5 orders of magnitude*.  Thus while the visual intricacy of the lattice pattern depends on colloidal plate size and numbers, the lattice character is preserved independent of plate size at a fixed area fraction near 17%.

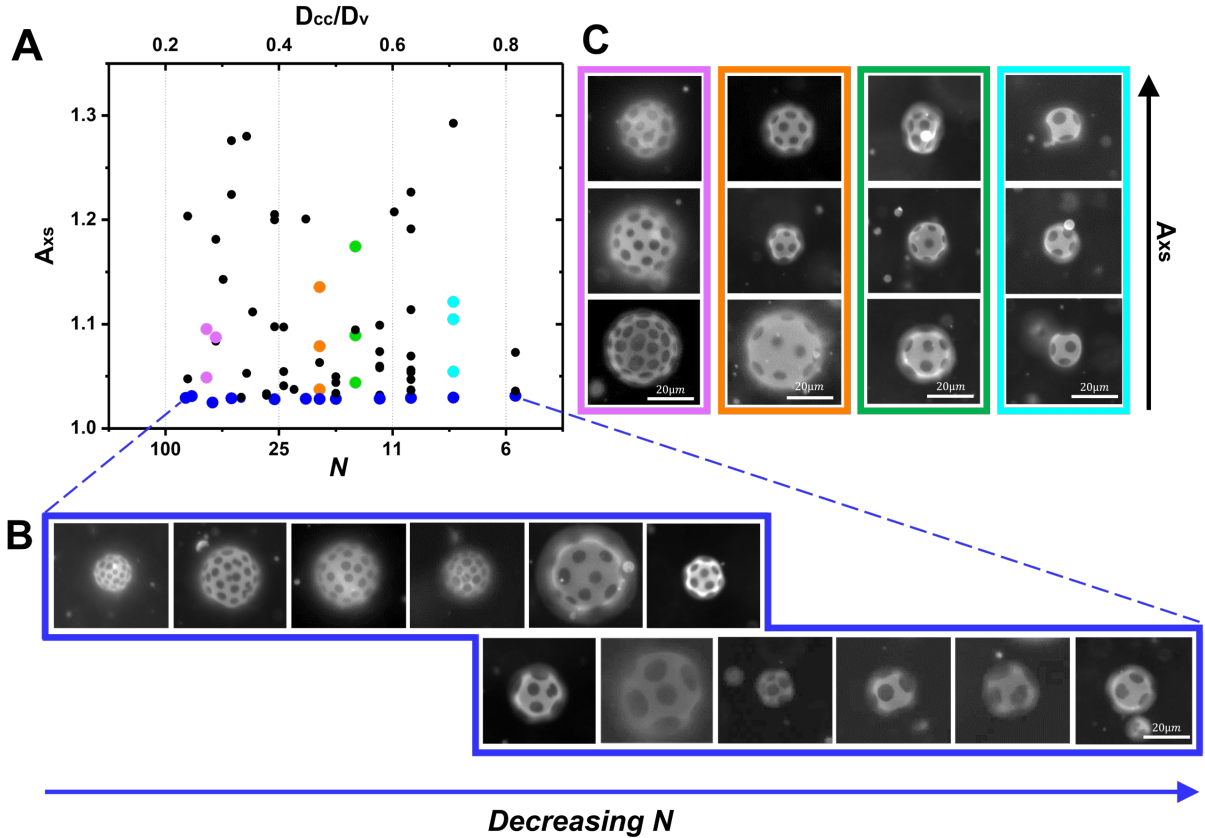

**Figure 3**. A) State space region for vesicle-encompassing quasi-lattice configurations, highlighting in blue, the vesicles with the smallest excess areas. These vesicles are shown in sequence in (B). C) Series of four vesicles each that, within each series, have similar domain number but varied excess areas. Each series is distinguished by colored points, highlighted in Part A.

Figure 3C emphasizes the minimal impact of increased excess area on overarching lattice structure, excerpting four series of different vesicles from the main plot in Figure 3A. Within each series, the different vesicles have a fixed colloidal plate number but vary $A_{xs}$. Thus it is seen that, at least near $\phi \sim 0.17$, for a wide range in $A_{xs}$, vesicle size, and plate numbers, a vesicle-encompassing quasi-lattice is a stable if not preferred configuration. Importantly Figures 3A and C demonstrate the robustness of the lattice structure to vesicle dehydration, for osmotic removal of at least up to 33% of the vesicle volume, producing $A_{xs} = 1.3$. Notably, with increased deflation, vesicles may sometimes deviate from a spherical shape in favor of an irregular ellipsoid but, more generally, distribution of undulations over the entire vesicle maintain an approximately spherical global shape.

*Cooperative Transition between Persistent Patterns*

Figure 4A expands the state space to include both types of persistent pattern classes: the vesicle-encompassing quasi-lattices and the closely associated configurations.  Independent of the numbers of colloidal plates per vesicle, the vesicle encompassing quasi-lattice is found only for $A_{xs}$ > 1.025, and the associated structures are found only for $A_{xs}$ < 1.015.  Thus there is a sharp horizontal boundary between the two persistent pattern types. The Supporting Information highlights how vesicles with dynamic randomly positioned plates overlap these two regions, but always within the range 1.01< $A_{xs}$ < 1.05.  Those vesicles containing dynamic plates may not yet have equilibrated or settled into low energy states, motivating focus on the vesicles with persistent arrangements.

Figure 4B shows the plate arrangements for pairs of vesicles highlighted in Figures 4A, having the same plate numbers on opposing sides of the boundary separating the regimes of quasi-lattice and associated pattern types.  Crossing the boundary, changing the excess area by about 1.5% profoundly alters the plate arrangement, with mostly linear configurations of closely associated colloidal plates at smaller $A_{xs}$ and the vesicle encompassing quasi-lattice at slightly greater $A_{xs}$. Either type of domain arrangement includes most if not all the plates in each vesicle. The sharpness of the state space boundary suggests a cooperative assembly or a phase transition. This dramatic change in configuration for a small change in $A_{xs}$ is a striking contrast to Figure 3 where large changes in excess area above $A_{xs}$ ~1.02 minimally affect the order of the vesicle-encompassing lattice arrangement.

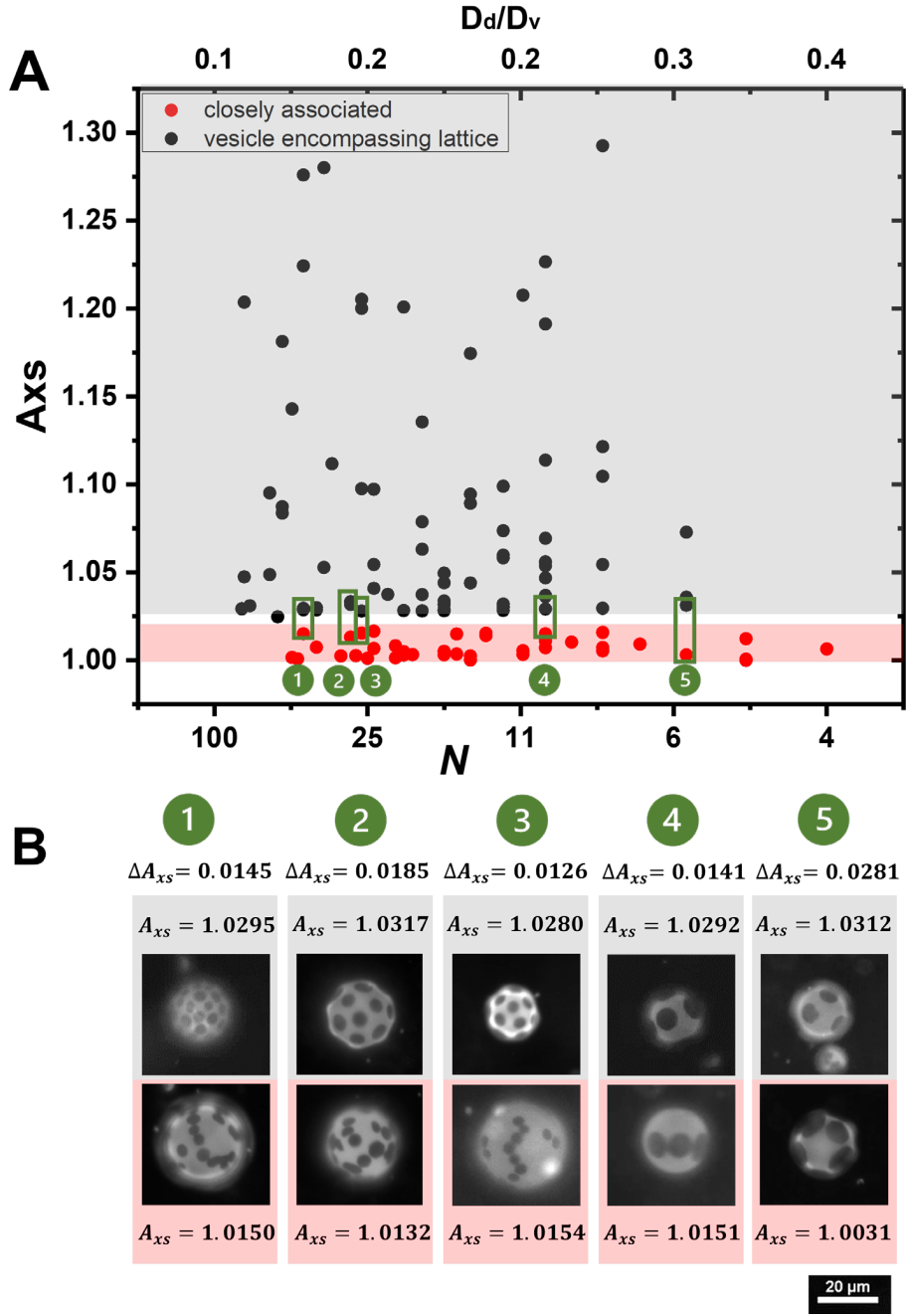

**Figure 4**. A) State space for vesicle-encompassing quasi-lattice in gray and closely associated configurations in red, with varied domain number (also translating to the domain to vesicle diameter ratio), and excess area. Five vesicle pairs, each traversing the boundary between regimes, are shown in (B), highlighting the sharp morphological change of the assembly for a small change in excess area across the state space boundary.

Changes in the plate-plate interaction with variations in $A_{xs}$ underlie the sharp transition in Figures 4A and B. This is evident in Figure 5A which summarizes the nearest neighbor edge-edge plate separations normalized on plate diameter, for vesicle-encompassing quasi-lattice and closely associated configurations. The plate separations are measured as the in-membrane distance between closest points on neighboring plates and averaged for plates towards the center of micrograph, where possible. Because the configurations are stable in time, the observed positioning provides a window into the effective plate-plate interaction.

Figure 5A establishes how plate separations in the closely associated configurations are finite and, for plates of a few microns, often the scale of a micron or less. Particularly striking is the uniformity in the separation of nearest closely associated domains within a given vesicle, seen in Figure 1C and in the small error bars of Figure 5A. This behavior suggests a short range repulsive core (still hundreds of nm) in the inter-plate potential and an intermediate attractive well, that is sufficiently deep to overcome thermal fluctuations. Additionally, the presence of a long range repulsion in the associated configuration is evidenced by the tendency for chained configurations.

By contrast, in the vesicle encompassing quasi-lattice, much larger plate-plate separations of several microns suggest repulsions with a range on the order of the plate size, up to 10 μm. Any attractive potential well present at high excess area is not accessible due to a substantial energy barrier. This is seen for $\phi = 0.17$ and, as expected, small variations in $\phi$ lead to small variations in the edge-edge separation, with slightly higher $\phi$-values giving slightly closer separations. Important to note is the near-complete collapse of data when scaled on the plate diameter suggesting that the colloid size presents an important length scale in these elasticity-based interactions.

The sharp change in edge-edge plate separation occurring at $1.015 < Axs < 1.025$ in Figure 5A for vesicles containing 4-100 domains is absent when vesicles contain only 2 domains. The data of Figure 5B, recalculated from our prior work[34] by scaling on $D_d$, reveal a more gradual shift in pairwise domain separation with $A_{xs}$. Indeed the cooperative multidomain transition in vesicles with several (4 of more) domains is not anticipated with vesicles containing only two domains at the same area fraction.

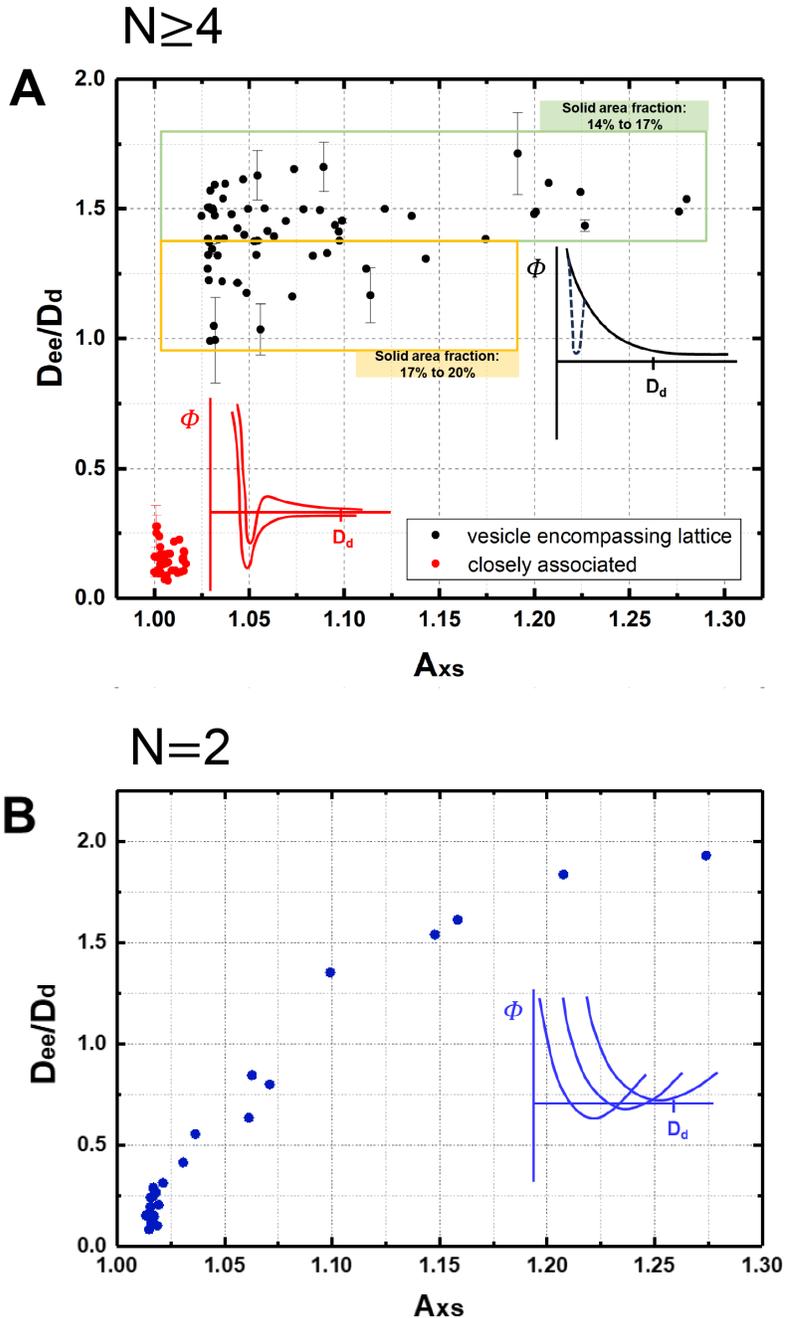

**Figure 5**. A) Normalized nearest neighbor separations, $D_{ee}/D_d$, measured from the edges of colloidal plates through the contoured fluid membrane at the closest points between pairs of plates and averaged, where possible, for each vesicle, then divided by the average domain diameter for that vesicle. Error bars reflect ability to resolve edge-edge separations and the impact of domain irregularity. Measurements were made within the limits of a given focal plane with resolution of ~0.5 μm, producing larger error bars for small vesicles and small domains. Error analysis is detailed in the Supporting Information. Variation in solid area fraction, indicated in (A) produces expected differences in the normalized domain separations in the vesicle encompassing quasi-lattice. B) Edge-edge separations recalculated from prior work[34] on vesicles with 2 domains per vesicle and a solid area fraction in the range 15%-19%.

**Discussion**

How elasticity mediates colloidal interactions in interfacial systems is a topic of ongoing research. While air-water and oil-water interfaces produce capillary repulsions between objects bordered by similarly curved menisci and attractions between objects bordered by menisci with opposing local curvature, such an effect is not likely at play in the current system, which is dominated by elastic interactions rather than capillarity. The current colloidal plates are all similar within each vesicle and, while the plates within some vesicles are more sharply faceted than plates in other vesicles, no inversion of curvature was evident in the fluid near the plates. Previous modeling[34] suggests that because a rigid solid pushes Gaussian curvature into the membrane fluid, and because this curvature tends to distribute analogous to a depletion potential to minimize bending costs, the fluid acts like a spring to push domains apart. With a reduction in excess area, the distribution of curved membrane between neighboring domains adjusts to produce an attraction at a finite distance.[24]

It is fascinating that, in Figure 5A, with multiple colloidal plates interacting within a single vesicle, attractive interactions develop and dominate configurations over an extremely small range in $A_{xs}$ as it is decreased toward unity and the vesicles inflate to approach a spherical shape. This scenario is reminiscent of the impact of reduced volume on the morphologies of lipids crystallizing in vesicles with different degrees of inflation.[35] Bending costs become sharply prohibitive for highly inflated vesicles, above $v_r \sim 0.98$, corresponding to $A_{xs}$ in the range 1.01 – 1.02, forcing the growing crystals to adopt petals that bend and wrap the vesicles. A potentially related wrapping behavior is seen here in the chains of plates that encircle vesicles at their widest parts, producing a composite shape that bends globally out of plane. This reduction in energy by the bending of an elongated structure, then, is one explanation for the preference for chained configurations. An important distinction between the chained structures observed here and chains of proteins[19,36] or nanoparticles[6,12] in certain experiments and simulations is that molecular curvature and interactions between proteins are key in the latter.[2,3,16,19] Such molecular interactions are absent here and consistent with the larger finite plate separation within chains of plates.

The ability to form a single chain is limited by the numbers and size of plates in each vesicle. With membrane domains approximated as flat circular plates and when plate number $N >$ $\pi {D_v}/{D_d}$, there are more plates than can encircle the vesicle in a single belt. This can occur at a fixed solid area fraction $\phi$ when many membrane domains nucleate in the initial formation of the colloidal plates. Then the additional plates must branch or form a second belt, or a plaque of closely associated or ordered plates may be preferred. Observations of locally ordered clusters of plates, with vacant fluid regions elsewhere in the membrane, as in the first micrograph of Figure 1C, suggest a reduction in the long range repulsion that favors chain formation.

The abrupt transition in configuration (Figure 4) and plate spacing (Figure 5A) with excess area for vesicles containing multiple colloidal plates presents a sharp contrast to the more gradual impact of excess area in vesicles containing only 2 plates in Figure 5B. It is unclear if the difference is simply that two-body interaction are always weaker when vesicle contain only two plates or if a first pair in vesicles with many domains shifts the entire interactive landscape. The considerably sharper change in plate separation with excess area in vesicles with many domains suggests a cooperative effect or phase transition.

This work focused on a single vesicle composition based on the lipid composition of 40/60 DPPC/DOPC molar ratio, giving a solid area fraction near 17%, with some variations based on temperature or loss of fluid membrane in a dislodged tether.  The composition was fortuitous in revealing the rich behavior reported here. One might naturally expect variations in solid area will shift the state space or phase diagram.  Exploring the impact of solid area fraction, a substantial undertaking, is an important future endeavor.

In a final note, we emphasize that multibody configurations seen here are long lived, suggesting at least metastability if not equilibrium. The lowest energy configurations for systems of plate-shaped colloids in 2D fluids will be further explored in future theoretical and experimental works.

**Summary and Outlook**

The elasticity-mediated interactions and organization of micron scale membrane inclusions holds great potential for a variety of applications requiring flexible material sheets.  These fluid-solid composites may someday provide critical functionality in applications requiring a variety of curved shapes, beyond the usual cylinders and cones of zero Gaussian curvature.  The current work demonstrates the robustness of the inclusion arrangements to deflation of giant vesicles, which imposes a non-spherical shape to the sheet.  The newly discovered sharp structural transition  between classes of configurations within a narrow window of state space, with just a percent change in inflation or excess area, opens the door to switchable optics that respond to the slightest touch.  Indeed the ability to achieve different patterns with a single material composition has important implications for using these strategies one step of the manufacturing sequence for patterned films.  At the same time this work provides insights into the interactions of rigid membrane domains, relevant to cell biology, separate from biomolecular and physio-chemical contributions.

**Materials and Methods**

*Materials*.  l,2-dipalmitoyl-*sn*-glycero-3-phosphocholine (DPPC) and 1,2-dioleoyl-*sn* -glycero-3-phosphocholine (DOPC) were purchased from Avanti Polar Lipids (catalog numbers 850375C and 850355C).  A tracer lipid, Rh-DOPE [1,2-dioleoyl-*sn*-glycero-3-phosphoethanolamine-N-(lissamine rhodamine B sulfonyl) ammonium salt, catalog number 810150C, was also purchased from Avanti Polar Lipids.

*Vesicles*. Giant unilamellar vesicles were electroformed on platinum wires following established procedures.[37,38]  Chloroform solutions of phospholipids in the desired proportions (40 mol% DPPC, 60 mol% DOPC), at an overall concentration near 25 mg/ml, and including no more than 0.2 lipid mol% (based on the lipids) of the Rh-DOPE tracer were deposited dropwise onto the wires and dried under nitrogen. The chamber was sealed between glass coverslips, the preheated 100 mOsm sucrose solution was introduced, and the chamber was maintained near 60°C as alternating current was applied at 3V and 10 Hz for an hour.  Electroforming at elevated temperature in the one phase region of the phase diagram ensured compositional uniformity of the membranes of all vesicles, while harvesting in a syringe brought the suspension to room temperature.  The 40/60 molar proportion of DPPC and DOPC produces vesicles with an expected solid domain area fraction of $17 \pm 3\%$, based on the phase diagram, with calculations in the Supporting Information.  Experimental sources error in achieving this targeted solid area fraction arise from temperature variations in the lab.

To produce vesicles containing relatively monodisperse domains, suspensions were reheated to 52°C and maintained at this temperature for 15 minutes to ensure complete melting of solid domains.  Subsequently, the suspension was cooled at a rate of 0.3°C/min to 42°C, still within the one phase region, and the temperature stabilized for 5 minutes.  Then the system was cooled to room temperature at rates in the range 0.25-1°C/min, with faster cooling rates selected to produce greater numbers of nucleated domains.[39] Different processing runs at these different cooling rates were needed to produce the range of domain numbers in this work, varying over an order of magnitude.  Experimental determination of the solid area fraction was done manually from epifluorescence micrographs, by determining the area of domains towards the center of the image (those best resolved) and multiplying the per-domain area by the numbers of domains.  Error arises from the resolution of domain edges, irregularities in domain shape, and modest domain polydispersity.

*Micropipette Manipulation*.  Excess area was assessed employing micropipette manipulation, with our apparatus described in detail previously.[40,41] Micropipettes were pulled on a Kopf Instruments micropipette puller and then forged on a Technical Products International microforge, to select the tip size and shape with nearly constant diameters in the range of 3-6 μm near the tip.  Suction was applied manually using a home-built suction manometer.

*Measurements*.  All vesicle images were obtained using a Nikon Diaphot TE-300 inverted fluorescence microscope.  Unless otherwise noted, measurements of plate and vesicle diameters were made manually using Nikon software, with a resolution as high as 1-2 pixels, at 0.34 um/pixel, unless otherwise noted.  Where possible measurements of domains and their separations were confined to the central region of vesicles, but a curvature correction in the Supporting Information was applied as needed.  Also, when vesicles had small excess areas, the membrane was taut, reducing the uncertainty in edge-edge separation measurements. At large excess areas, edge-edge separation between plates, even towards a the center of vesicle image, carried error due to a component of the membrane contour perpendicular to the viewing plane. These edge-edge values were corrected also as described in the Supporting Information.

Following electroforming and reprocessing in 100 mOsm sucrose, vesicles were osmotically conditioned to produce different patterns. In the studies of Figure 1, stock vesicle solutions reprocessed in 100 mOsm sucrose were diluted by half in DI water producing 50 mOsm sucrose on the vesicle exterior. Water driven into the vesicles in these studies caused the membrane to become tense. Vesicles were imaged after 15-20 minutes, before there was substantial vesicle bursting. Control studied employed dilutions of the original stock suspension in glucose solution, to give a 50-50 glucose-sucrose solution on the vesicle exterior with neutral osmotic conditions of 100 mOsm.

In the studies of Figures 2-5, stock vesicle suspensions were mixed in a 1:10 ratio with sucrose solutions having concentrations in the range 50-120 mOsm, to produce varying degrees of vesicle inflation or deflation. Vesicles were imaged after equilibration, 15 minutes or longer, in an open chamber that allowed micropipette manipulation. Excess area measurements were conducted by aspirating the vesicle gently into a pipette and then reducing the suction to hold the vesicle in the micropipette without stretching the membrane. From an image at this point, the vesicle area and volume were quantified (detailed in the Supporting information), and excess area calculated via equation 2. It should be noted that such aspiration, especially when vesicles was first drawn into the pipette, tended to disrupt any patterning. Images of patterns were recorded prior to the aspiration of a given each vesicle studied.

*Supplementary Information:*
1. Calculation of solid fraction from phase diagram,
2. Delaunay triangulation and images,
3. Calculation of excess area and error analysis
4. Center-center and edge-edge domain spacing for vesicle encompassing lattice and error
5. State space including vesicles with dynamic disordered domains.

*Acknowledgements*. This work was made possible by support from the DOE, DE-SC0017870. We are grateful to Greg Grason for numerous discussions on this work and his specific input on the Lindemann criterion.

*Author Contributions*: MMS conceived and developed the experimental program performed design calculations and error analysis, reviewed data, and conceptualized data analysis and interpretation. WX conducted all experiments and data analysis. MMS wrote this manuscript and WX developed all figures.

*Supplemental Information*

**Bending Energy-Driven Cooperative Patterning of 2D Colloids in Elastic 2D Fluids**

Weiyue Xin,[1] and Maria M. Santore[2]*

1. Department of Chemical Engineering
   University of Massachusetts Amherst

2. Department of Polymer Science and Engineering
   University of Massachusetts Amherst
   120 Governors Drive
   Amherst, MA 01003
   santore@mail.pse.umass.edu

# 1. Estimating the solid area fraction from the phase diagram

We consider a membrane consisting of one mole total of DPPC and DOPC lipids. The membrane partitions into fluid (liquid) and solid domains containing $L$, and $S$ moles, respectively, such that $L + S = 1$. That is, $L$ is the moles of DOPC and DPPC taken together in the fluid membrane phase ,and $S$ is the moles of DOPC and DPPC, taken tighter, in the solid membrane phase.

The overall mole fraction of DPPC in the entire vesicle is $z_{DPPC}$, and the mole fractions of DPPC in the liquid and solid phases are $x_{DPPC}$ and $y_{DPPC}$, respectively.

A mass balance on the DPPC can be rearranged to give the form of the inverse lever arm rule for the amount of solid, given the composition, $x_{DPPC}$, $y_{DPPC}$, and $z_{DPPC}$ on a tie line:

$$S = \frac{z_{DPPC} - x_{DPPC}}{y_{DPPC} - x_{DPPC}} \qquad (1)$$

In equation 1, the solid is quantified in terms of $S$, the molar proportion of solid relative to the total moles in the entire membrane.

In order to obtain the solid area fraction, $\phi$, one must include the molar areas of the fluid and solid domains, $\underline{A}_L$ and $\underline{A}_S$, respectively. Their ratio, $R$, is $R = \underline{A}_S/\underline{A}_L$. Then

$$\phi = \left[1 + \frac{1}{R} \frac{(y_{DPPC} - z_{DPPC})}{(z_{DPPC} - x_{DPPC})}\right]^{-1} \qquad (2)$$

At room temperature, the molecular area of the fluid is approximated by the area of pure DOPC in the $L_\alpha$ phase, since the fluid membrane is mostly DOPC. The molecular area of DOPC in the $L_\alpha$ phase at 30°C is reported as 72.5A.[1] Then, using a coefficient of thermal expansion of 0.003/°C also based on the work of Nagel and Tristram-Nagel,[1] the molecular area at room temperature is found: $A_L(\text{DOPC @ 22°C}) = A_L(\text{DOPC@ 30°C}) \, e^{\kappa \Delta T} = 70.8 \text{ Å}^2$.

The solid molecular area is estimated based on the $P_{\beta'}$ phase of DPPC at room temperature,[2,3] which is established to be nearly pure[4,5] such that $y_{DPPC} \sim 1$. The area of a DPPC molecule in the solid phase is estimated by starting with a value measured at 50°C in the $L_\alpha$ phase, $A_L(\text{DPPC @ 50°C}) = 64.0 \text{ Å}^2$.[1] From here, the footprint of DPPC in solid phase at room temperature is calculated in three steps: cooling from 50°C to $T_m \sim 42$°C using a coefficient of thermal expansion in the range $\kappa = 0.003 - 0.006$ /°C;[6] a 17% reduction in area for the phase transition from $L_\alpha$ to $P_{\beta'}$ at 41-42°C;[7] and cooling of the solid to 22°C using a coefficient of thermal expansion in the range $\kappa = 0.003 - 0.006$ /°C. [6] This gives a range for the solid molecular area of $A_S$(DPPC@ 22°C) = 45.2 – 49.0 Å².

Finally in applying equation 2 to estimate $\phi$, a composition of $x_L$(DPPC) = 0.18, was estimated for the end of the room temperature tie line, based on measurements of the liquidus curve of the phase diagram, by cooling vesicles having 20 mol% DPPC, and noting the temperature where solid domains first appeared.

Application of equation 2 therefore gives, for membranes containing 40 mol% DPPC, a solid fraction of:

$$\Phi = (1 + \frac{70.8}{47} * \frac{0.4 - 1.0}{0.18 - 0.4})^{-1} = 0.19$$

By considering different estimates for the coefficient of thermal expansion and compositions at the ends of the tie line we arrive at $\phi = 0.19 \pm 0.03$.

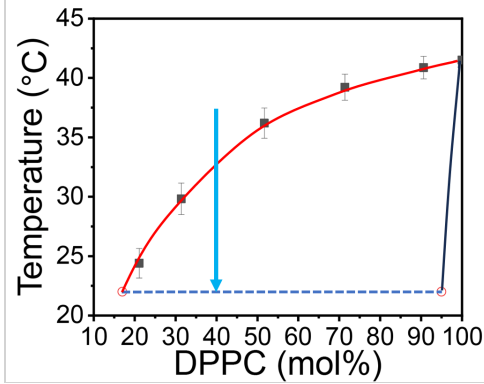

**Figure S1**. Phase Diagram for DOPC/DPPC mixed membranes, reproduced from previous work our this lab.[8]

**2. Limited Delaunay Triangulation.**

A limited version of Delaunay Triangulation, a type of pattern analysis, was pursued as a means to validate the by-eye categorization of colloidal patterns into the three types. The term "limited" is employed because a more detailed and complete use of the method, taking into account the curvature of the vesicle and including all domains in 3D, might provide further information into the nature of patterning of closed or sphere-like objects. The current work, however, cannot identify all domains on each vesicle and does not attempt geometrical transformations to account for the pseudo-spherical vesicle contour. By selecting only a subset of domains on each vesicle, the "limited" method avoids much image analysis error, at the expense of rigorous pattern analysis. The current exercise using Delaunay triangulation in a limited fashion simply aims to demonstrate that the by-eye classification of patterned vesicles can be done without bias.

Delaunay triangulation is a method in computational geometry that connects a given set of points on a plane into triangles, chosen so as to maximize the minimum angle of all the triangles in the triangulation and to allow no point in the dataset to exist inside the circumcircle of any triangle. This means that in the Delaunay triangulation, not every pair of points will be directly connected, and the edges of each triangle do not intersect.

The analysis of each vesicle commenced with the measurement of the vesicle diameter from an equatorial view and, from a top or bottom view, the identification of individual dark domains and their center points. Next, choosing the domain closest to the center of the vesicle by eye, the six closest domains to the first were identified. Then, the Delaunay-Voronoi plugin available in the FIJI software was utilized to construct the triangulation, with the triangle sides connecting the domain centers. Then, FIJI calculated the lengths of the sides all the triangles, which provided a list of center-center domain distances based on the central and up to 6 neighboring domains.

With the idea that a regular grid of domains would have nearly identical domain spacings but elaborate patterns might be evident in other domain spacings, we calculated a coefficient of variation, a $CV$, taking the standard deviation of the domain spacings on a given vesicle image and normalizing by the mean.

$$CV = \frac{standard\ deviation\ of\ center-center\ spacing}{Mean\ center-center\ domain\ spacing}$$

For a perfectly ordered system, all the domains would be regularly spaced and the standard deviation and therefore the $CV$ would vanish. An irregular grid would be expected to have a finite but small $CV$, and different patterns might have different $CV$ s.

Detailed examples of Delaunay triangulation patterns are shown in Figure S2 for arbitrary domain arrays while Figure S3 shows closeups of the triangulations from Figure 1 A-C.

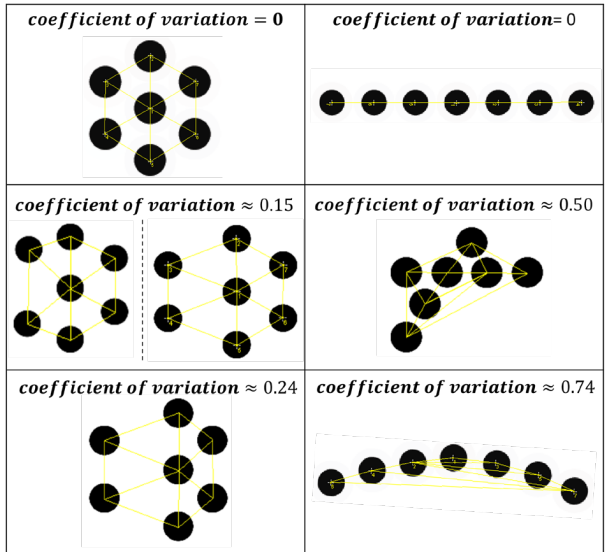

**Figure S2.** Example domain arrangements, their triangulation patters and coefficients of variation.

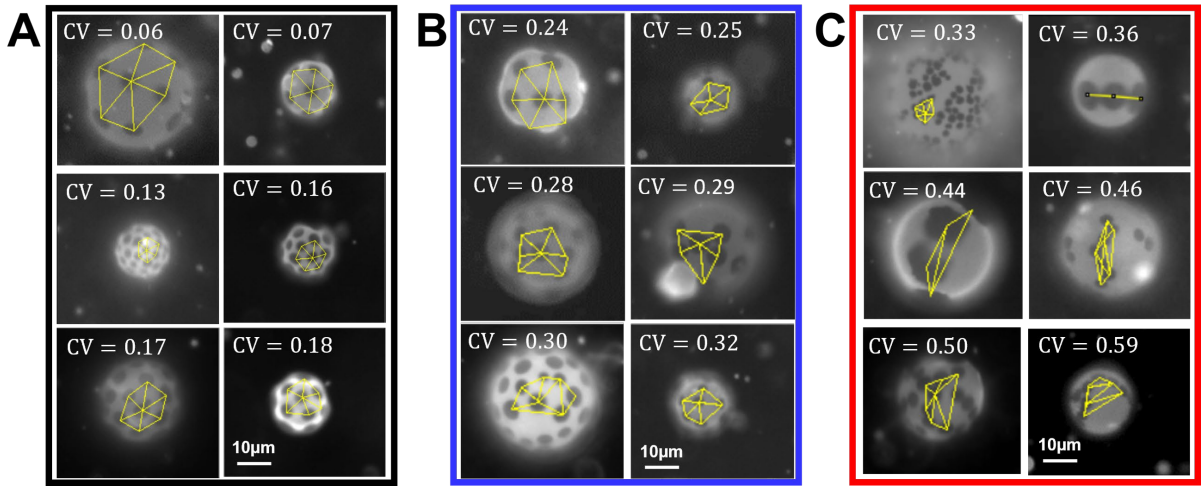

**Figure S3.** Enlargement of insets from Figure 1A-C showing Delaunay triangulation for example patterns on vesicles. A) vesicle-encompassing lattices B) dynamic disorder and C) closely associated.

The error in Coefficient of Variation (*CV*) can be attributed to two main factors:

Positions of domain centers: Measurement error or resolution constraints of ~2 pixels in the position of the domain center corresponds to as much as 15% error in the calculation of a single center-center domain separation.  Typically the error is much less than this, just a few percent, for large domains and large absolute separations.  The 15% worst case scenario occurs on small vesicles with many domains where the triangle edges are short (~5 μm) and push the limits of our resolution.  However, because the lengths of all the triangle sides area not independent, the *average* separation length, for instance for a regular array, exhibits far less error than 15%, even for the worst case. Mis-identifying the center positions more directly impacts *CV*, by increasing the standard deviations of the triangular edges resulting from a pattern on small vesicles with many domains.  This is seen, in a worst case way, by the calculated *CV* values for the hex arrays in the above table.  Even with this error in *CV*, the clustering of *CV* by pattern type is clear and outside this error.

2D Projection of 3D Structures: The *CV* values ascertained by FIJI are based on 2D micrographs and do not correct for the curvature of pseudo-spherical vesicles. The impact on the measured center-center distance is small for domain pairs towards the vesicle center but more significant when the line between the centers of a domain pair has a substantial radial component.  Then the radial component of the center-center line appears shorter, and domains are reported as being closer than they really are.  This error does not occur for the theta component of center-center vectors between neighboring domains.  For instance, the actual arc length along the membrane in the radial direction is $R \sin^{-1} \frac{mesaured\ radial\ component}{R}$ . The worst case for this type of error therefore occurs when the line connecting neighboring domains is entirely radial, with no theta component. Then, for instance, if the center-center line starts at the middle of the vesicle and ends three quarter the distance from the center to the edge image, then center-center domain distance will be appear to be about 88% of the actual in-membrane contour between the two centers. For this reason, Delaunay analysis employed domains positioned more towards the centers of vesicles, so that error was only a few % in many cases.

### 3. Micropipette Measurements to Determine $A_{xs}$

The excess area, $A_{xs}$, is a dimensionless measure of the vesicle membrane area relative to that needed to encapsulate the vesicle's contents as a sphere. Increased excess area allows the vesicle to take on a variety of shapes (without stretching), or for the membrane to undulate and bend locally. $A_{xs}$ is calculated by dividing the actual membrane area of the vesicle, $A$, by the area $A'$ required to envelop the vesicle's contents in a spherical shape. When excess area equals 1, the vesicle is a sphere. Any other shape will produce an excess area exceeding 1.

For a vesicle of volume $V$, the radius of an equivalent sphere is

$$r' = \left(\frac{3V}{4\pi}\right)^{\frac{1}{3}} \quad . \tag{3}$$

Then the equivalent area needed to encapsulate this volume is

$$A' = 4\pi r'^2 = 4\pi * \left(\frac{3V}{4\pi}\right)^{\frac{2}{3}} \tag{4}$$

The excess area is the ratio of the actual vesicle area

$$A_{xs} = \frac{A}{A'} = \frac{A}{(4\pi)^{\frac{1}{3}} * (3V)^{\frac{2}{3}}} \tag{5}$$

Micropipette aspiration at low suction pressure to ensure membrane strain below 0.1%, was utilized to measure $A$ and $V$, and then $A_{xs}$ was calculated for each vesicle from equation 5, which corresponds to equation 2 of the main paper.

A partially-deflated vesicle, when aspirated into a micropipette forms a predominantly cylindrical projection within the micropipette and a bulbous shape outside of it. The bulb outside the micropipette takes a spherical shape except for flat solid domains. For vesicles containing ~ten to a hundred small domains, the membrane often approaches a uniform distribution of domains between the portion of the vesicle inside and external to the pipette. For vesicles containing 2-5 relatively large domains, only the fluid membrane is typically drawn into the micropipette.

Depending on a vesicle's excess area, there may (in Figure S4A) or may not be (in Figure S4B) sufficient membrane to fully establish a projection inside the micropipette at low suction. This distinction leads to the different analysis approaches, below, to estimate the vesicles' area and volume from microvideo images.

After vesicle aspiration, we measured two parameters: $D_{asp}$, the diameter of the vesicle outside the pipette (neglecting the impact of solid domains on its otherwise spherical contour), and $L$, the entire length of the projection inside the micropipette, including its cap. The pipette's inner diameter near the tip, $D_p$, is measured in advance. From these values, we calculate $A$ and $V$. For vesicles with solid domains, we also measure their diameters, denoted $D_d$.

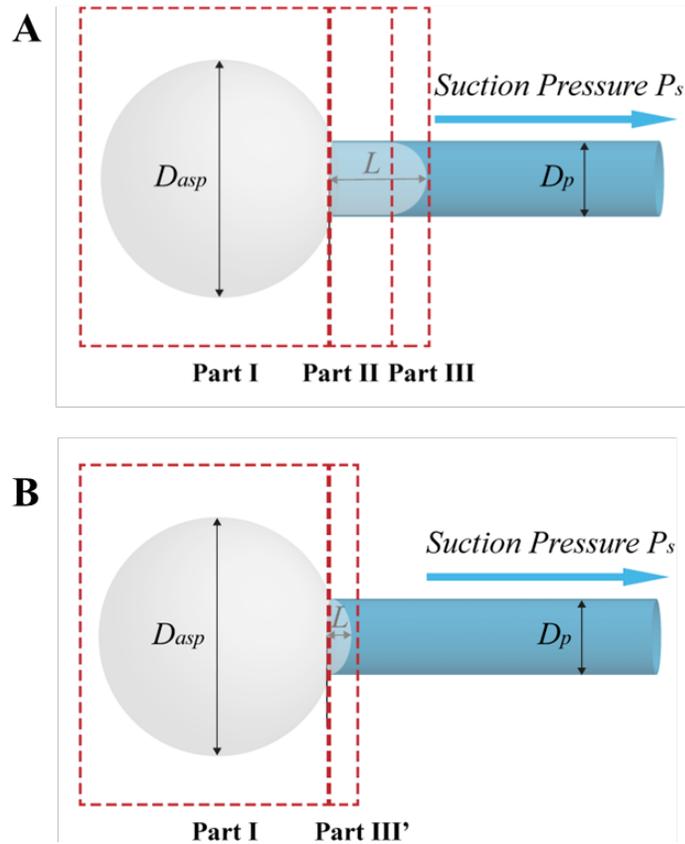

**Figure S4.** Shapes of vesicles aspirated into a micropipette for A) deflated vesicle producing a complete projection, and B) a vesicle with small excess area, barely entering the micropipette

Calculation of $A$ and $V$ is accomplished by considering parts of the vesicle which are shaped either as truncated spheres, Parts $I$, $III$, and $III'$, or as a cylinder, Part $II$. When vesicles have substantial projections, then $A = A_I + A_{II} + A_{III}$  and $V = V_I + V_{II} + V_{III}$. When the vesicle is minimally aspirated such that $L < D_P/2$ then $A = A_I + A_{III'}$,  and $V = V_I + V_{III'}$. Formulas for the various vesicle parts are given in Table S1.

**Table S1: Expressions for Areas and Volumes of Vesicle Parts**

| | |
|---|---|
| 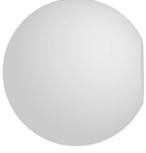 Part I | $A_I = \pi D_{asp}{}^2 - \pi D_{asp} * h$ <br><br> $V_I = \dfrac{1}{6} \pi D_{asp}{}^3 - \pi h^2 (\dfrac{D_{asp}}{2} - \dfrac{h}{3})$ <br><br> $h = \dfrac{D_{asp}}{2} - \dfrac{1}{2}\sqrt{D_{asp}{}^2 - D_p{}^2}$ |
| 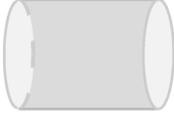 Part II $\left( L > \dfrac{D_p}{2} \right)$ | $A_{II} = \pi D_p * (L - \dfrac{D_p}{2})$ <br><br> $V_{II} = \dfrac{1}{4} \pi D_p{}^2 * (L - \dfrac{D_p}{2})$ |
| 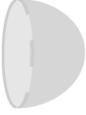 Part III $\left( L > \dfrac{D_p}{2} \right)$ | $A_{III} = \dfrac{1}{2} \pi D_p{}^2$ <br><br> $V_{III} = \dfrac{1}{12} \pi D_p{}^3$ |
| 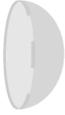 Part III' $\left( L \leq \dfrac{D_p}{2} \right)$ | $A_{III'} = 2\pi * \left( \dfrac{L}{2} + \dfrac{D_p{}^2}{8L} \right) * L$ <br><br> $V_{III'} = \pi L^2 * \left( \dfrac{L}{6} + \dfrac{D_p{}^2}{8L} \right)$ |

**Error Analysis:**

Error in determination of the excess area can arise from the of resolution in the measured parameters, $D_{asp}$, $L$, and $D_p$, up to ~ 2 pixels. This can produce up to 2.5% error in the determination of $A_{xs}$, with the worst case for small vesicles and large diameter micropipettes, for instance a 15um diameter vesicles and a 6 μm micropipette. Working at low suction maintains the areal strain of the membrane below 0.1%, another source of error which is extremely small, and systematic across all measurements. Finally, The solid domains exist as flat plates integrated in the fluid membrane, producing deviations between the spherical treatment of the vesicle bulb outside the pipette its actual shape. Solid domains within the projection tend to bend cylindrically, not altering the projection shape and not contributing to error. The error from the flatness of the colloidal plates on the spherical bulb region depends on their number, and becomes extremely small for large N. Since the area and volume of the bulb part of the aspirated vesicle are a variable portion of the overall area and volume, the impact of error in the area and volume in the bulb on the overall error is variable, and sets an upper limit on this type of error, shown in Figure S5.

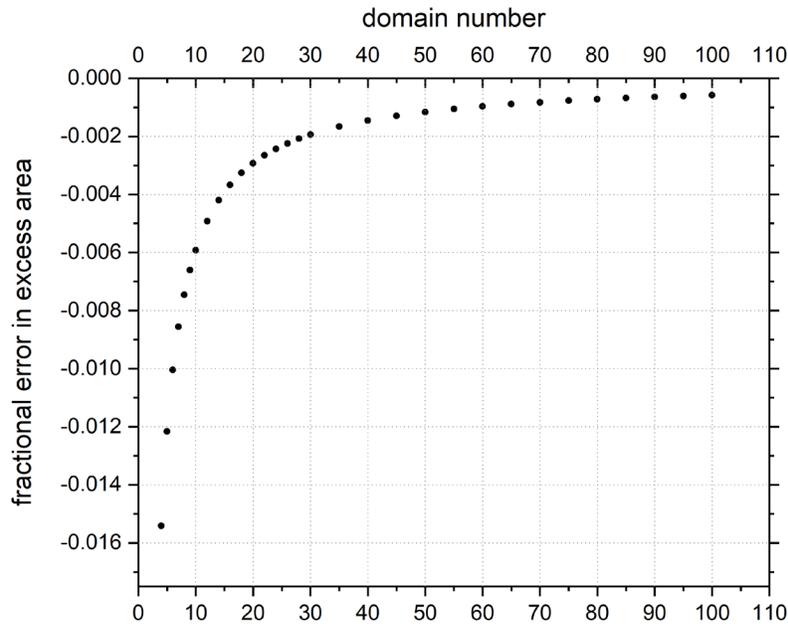

**Figure S5.** Fractional error in $A_{xs}$ (an upper limit) resulting from the treatment of the bulb portion of the aspirated vesicle as a sphere, neglecting the flat plate domains. The calculation assumes aa fixed solid area fraction of 17% as $N$ changes.

**4. Center-center and edge-edge domain spacing for vesicle encompassing lattice**

When $N$ domains are roughly evenly distributed on a sphere, it is possible to estimate the center-center domain spacing, $D_{cc}$, by dividing the entire sphere surface into $N$ sections, in Figure S6.

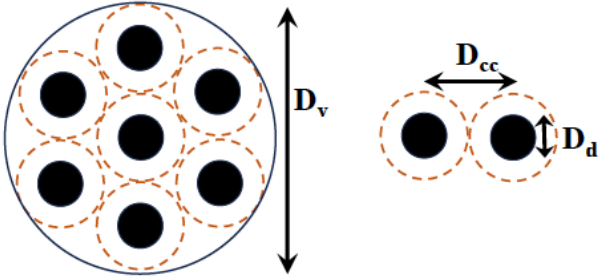

**Figure S6.** Schematic of domains on a vesicle, vesicle diameter, domain diameter, and center-center domain spacing. The domains at the edge of the image area not shown in the schematic. In experiments they may or may not be visible depending on focal plane, vesicle size, and depth of field. In experiments, $D_v$ is always measured in an equatorial view.

Each section is then approximated as a circle whose diameter equals the center-center domain spacing. This gives

$$\frac{D_{cc}}{D_v} = \sqrt{\frac{4}{N}} \qquad (6)$$

This relationship was tested by measuring the domain spacing of 78 vesicles that were classified as containing vesicle encompassing lattices. The results, in Figure S7A show excellent agreement between experiments and equation 6 while Figure S7B confirms the inverse half scaling with $N$.

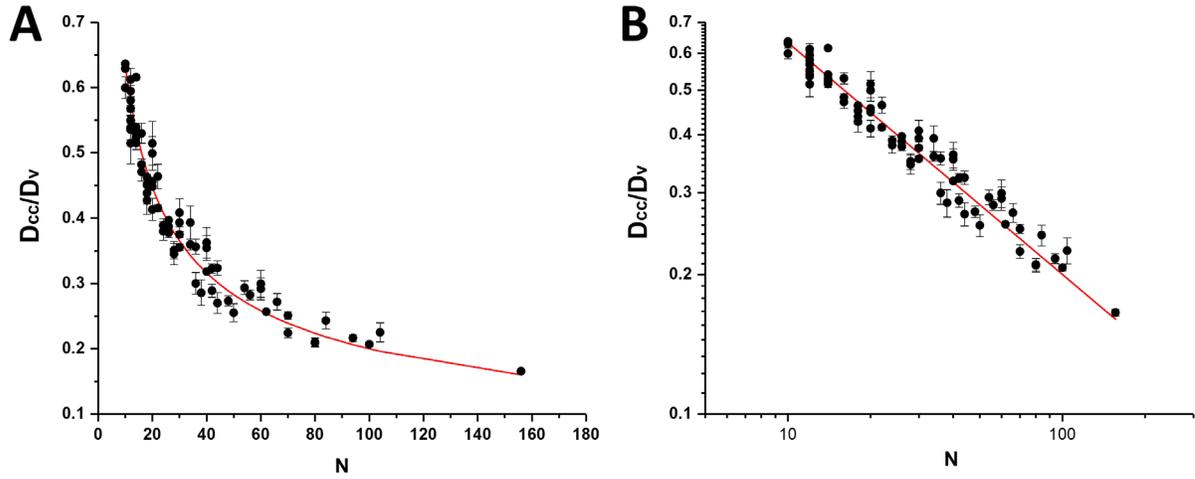

**Figure S7.** A. For vesicle encompassing lattices, the center-center distance measured on micrographs compared with equation 6 in red. B. Same data on log scale reveals -1/2 power scaling (red line slope.)

Figure 7A establishes that equation 6 reasonably describes normalized center-center domain spacings in the vesicle-encompassing lattice, independent of any vesicle deflation, i.e. independent of increased excess area. This is expected because if modest deflation enables membrane fluctuations, then the vesicle can appear spherical with measured vesicle diameter $D_v$', somewhat smaller than the fully inflated value. Then with $N$ domains each assigned a circular membrane region, which also appears smaller than at full inflation, the center-center separation, normalized on $D_v$', will still follow equation 6.

Different from the center-center distances, however, the edge-edge domain separation will become disproportionately smaller with vesicle deflation. Here, the apparent and actual vesicle diameters are related by an approximate relationship: $D_v \big/ D'_v \sim \sqrt{A_{xs}}$. While the presence of excess area is often evident in the equatorial view, represented schematically in Figure S8, there is extra membrane between domains visualized in the top or bottom focus, even towards the center of the image where curvature in the radial direction is expected to be small. Quite often the membrane can appear bright as a result of its bending fluctuations, in addition to the curvature that is evident in the equatorial view.

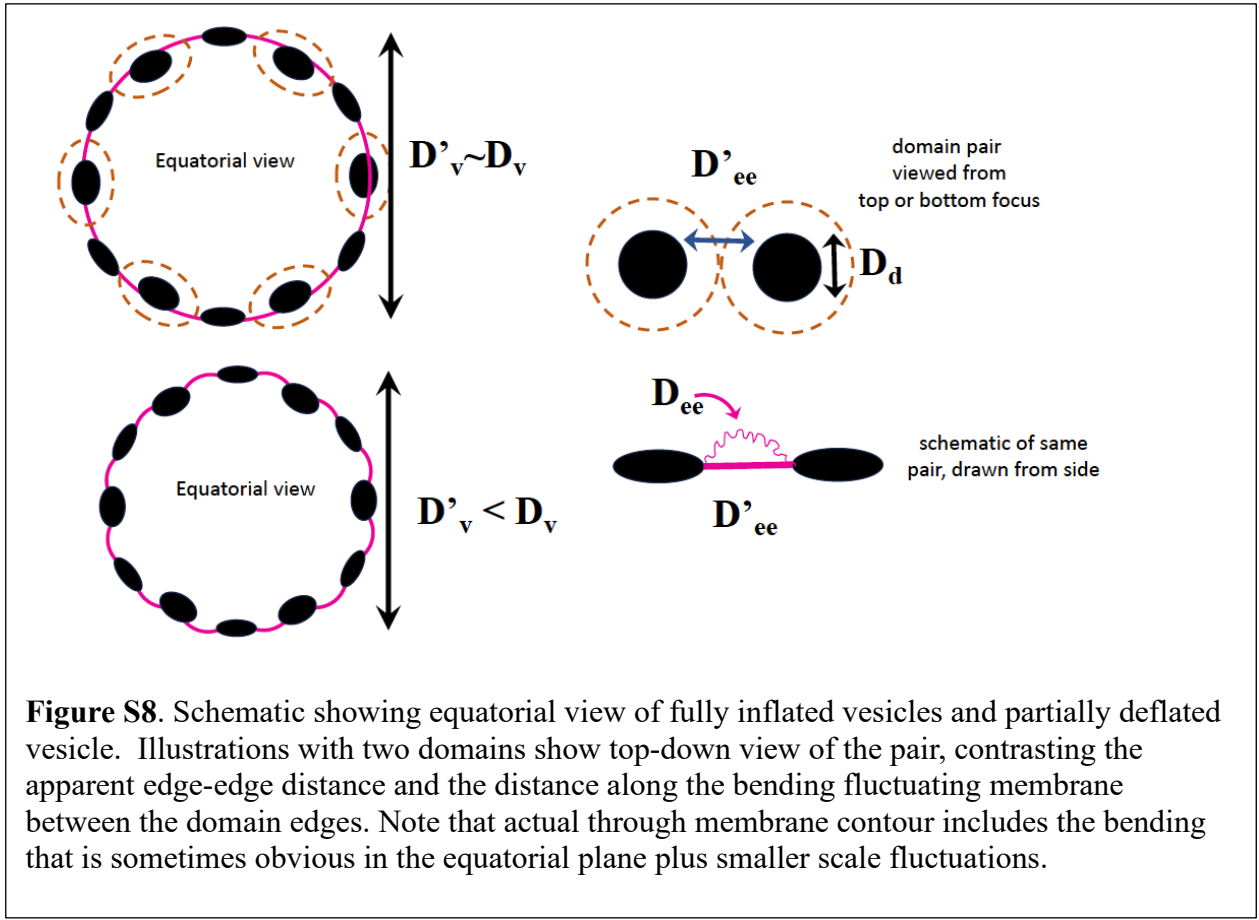

**Figure S8**. Schematic showing equatorial view of fully inflated vesicles and partially deflated vesicle. Illustrations with two domains show top-down view of the pair, contrasting the apparent edge-edge distance and the distance along the bending fluctuating membrane between the domain edges. Note that actual through membrane contour includes the bending that is sometimes obvious in the equatorial plane plus smaller scale fluctuations.

When we estimate the edge-edge domain separation through the membrane fluid phase the membrane fluctuations can be accounted for in an approximate fashion when the $A_{xs}$ is quantified separately with micropipettes.

We can write for the actual fluid membrane along the circumference:

$$\pi D_v - N^{1/2} D_d. \tag{7}$$

Here $D_v$ is the diameter of the fully inflated sphere and an approximation for the solid portion of the circumference is subtracted. Then the actual <u>edge-edge</u> distance between distributed domains is (dividing by $N^{1/2}$ domains in the equatorial slice)

$$D_{ee} = \frac{\pi D_v}{N^{1/2}} - D_d \tag{8}$$

And using $D_v'$ for the apparent (measured) diameter

$$D_{ee} = \frac{\pi D'_v \sqrt{A_{xs}}}{N^{1/2}} - D_d \tag{9}$$

However, $D'_{ee}$ is the value that is measured in a micrograph focused on the top or bottom vesicle view:

$$D'_{ee} = \frac{\pi D'_v}{N^{1/2}} - D_d \tag{10}$$

And so the apparent edge-edge separation can be corrected to include the excess membrane that bends between neighboring domains:

$$\frac{D_{ee} + D_d}{D'_{ee} + D_d} = \sqrt{A_{xs}} \tag{11}$$

Or

$$D_{ee} = \sqrt{A_{xs}} \left[ D'_{ee} + D_d \right] - D_d \tag{12}$$

Or

$$\frac{D_{ee}}{D_v} = \sqrt{A_{xs}} \left[ \frac{D'_{ee}}{D_v} - 1 \right] + 1$$

Note that when the excess area approaches unity, $D_{ee}$ approaches $D'_{ee}$. For an apparent edge-edge separation of $\frac{D'_{ee}}{D_v} = 1.4$ the actual through-membrane separation can be as large as $\frac{D_{ee}}{D_v} =$ 1.456 for large excess areas of 1.3, a worst case estimate. For a closer apparent edge-edge separation of separation $\frac{D'_{ee}}{D_v} = 1.4$, and a large excess areas of 1.3, $\frac{D_{ee}}{D_v} = 1.342$, with a slightly smaller area. Therefore the effect of excess area, domain irregularities, and imaging resolution contribute to the error in Figure 5A of the main paper. The latter two effects are more important than the impact of excess area.

## 5. State space including vesicles with dynamic disordered domains.

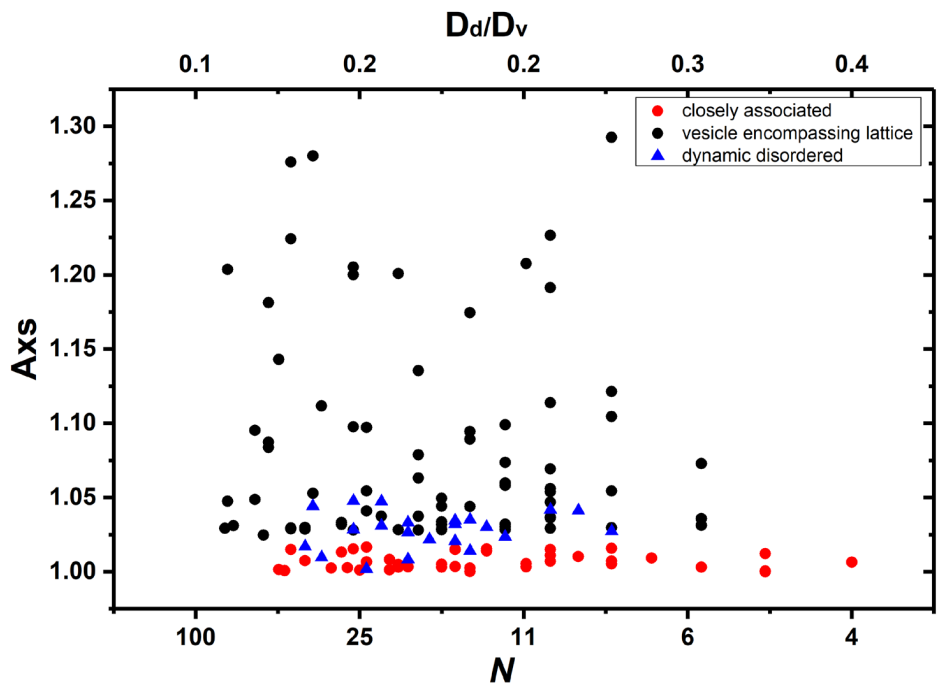

**Figure S9**. State space including data (blue triangles) for vesicles with dynamic disordered patterns. Vesicles having persistent domain arrangements are represented by black circles (vesicle-encompassing lattice) and red circles (closely associated configurations.)